\documentclass[twocolumn,floatfix,showpacs,prb]{revtex4}
\usepackage{graphicx}
\usepackage{amsmath}
\usepackage{amssymb}
\usepackage{bm}

\begin{document}

\title{Topological quantum number and critical exponent from conductance fluctuations at the quantum Hall plateau transition}
\author{I. C. Fulga}
\affiliation{Instituut-Lorentz, Universiteit Leiden, P.O. Box 9506, 2300 RA Leiden, The Netherlands}
\author{F. Hassler}
\affiliation{Instituut-Lorentz, Universiteit Leiden, P.O. Box 9506, 2300 RA Leiden, The Netherlands}
\author{A. R. Akhmerov}
\affiliation{Instituut-Lorentz, Universiteit Leiden, P.O. Box 9506, 2300 RA Leiden, The Netherlands}
\author{C. W. J. Beenakker}
\affiliation{Instituut-Lorentz, Universiteit Leiden, P.O. Box 9506, 2300 RA Leiden, The Netherlands}
\date{October 2011}
\begin{abstract}
The conductance of a two-dimensional electron gas at the transition from one quantum Hall plateau to the next has sample-specific fluctuations as a function of magnetic field and Fermi energy. Here we identify a universal feature of these mesoscopic fluctuations in a Corbino geometry: The amplitude of the magnetoconductance oscillations has an $e^{2}/h$ resonance in the transition region, signaling a change in the topological quantum number of the insulating bulk. This resonance provides a \textit{signed} scaling variable for the critical exponent of the phase transition (distinct from existing positive definite scaling variables). 
\end{abstract}
\maketitle

\section{Introduction}
\label{sec:intro}

A two-dimensional electron gas in a strong perpendicular magnetic field has an insulating bulk and a conducting edge. While the conductance $G_{\rm edge}$ for transport along the edge is quantized in units of $e^{2}/h$, the conductance $G_{\rm bulk}$ for transport between opposite edges is strongly suppressed. This is the regime of the (integer) quantum Hall effect.\cite{QHE86} Upon variation of magnetic field or Fermi energy, a transition occurs in which $G_{\rm edge}$ varies from one quantized plateau to the next, accompanied by a peak in $G_{\rm bulk}$.

The quantum Hall plateau transition is not smooth, but exhibits fluctuations reminiscent of the universal conductance fluctuations in metals.\cite{Sim91,Mai94,Byc96,Cob96,Mac01,Hoh02,Pel03} In small samples the fluctuations take the form of sharp peaks, due to resonant scattering between opposite edges mediated by quasi-bound states in the bulk.\cite{Jai88} In larger samples the intermediate states form a percolating network at the plateau transition, resulting in a smooth peak in $G_{\rm bulk}$ with rapid fluctuations superimposed.\cite{Coo97}

The conductance fluctuations are typically analyzed in a context that emphasises their random, sample-specific nature.\cite{Hoh02} Here we wish to point out one feature of these fluctuations that has a sample-independent \textit{topological} origin. 

The quantum Hall plateau transition is a topological phase transition, because a topological quantum number ${\cal Q}$ changes from one plateau to the next.\cite{Tho82} We show that a change in ${\cal Q}$ is associated with a resonance in the amplitude $\Delta G$ of the magnetoconductance oscillations in a ring (Corbino) geometry. Each unit increment of ${\cal Q}$ corresponds to a resonant amplitude $\Delta G=e^{2}/h$ (times spin degeneracies). Our analysis relies on a scattering formula for the topological quantum number, which relates ${\cal Q}$ to the winding number of the determinant of the reflection matrix.\cite{Bra09,Ful11} 

The analytical considerations in Sec.\ \ref{chern} are supported by numerical calculations in Sec.\ \ref{xp}. These show, in particular, that the plateau transition in a disordered system is reentrant: there are multiple increments of ${\cal Q}$, alternatingly $+1$ and $-1$, each associated with an $e^{2}/h$ resonance in $\Delta G$. We conclude in Sec.\ \ref{conclude} by showing that the critical exponent of the phase transition can be extracted from the sample-size dependence of the width of the resonance, in a way which preserves information on which side of the phase transition one is located. The Appendices contain details of the numerical calculations.

\section{Topological quantum number and conductance resonance}
\label{chern}

\begin{figure}
\includegraphics[width = 0.6\linewidth]{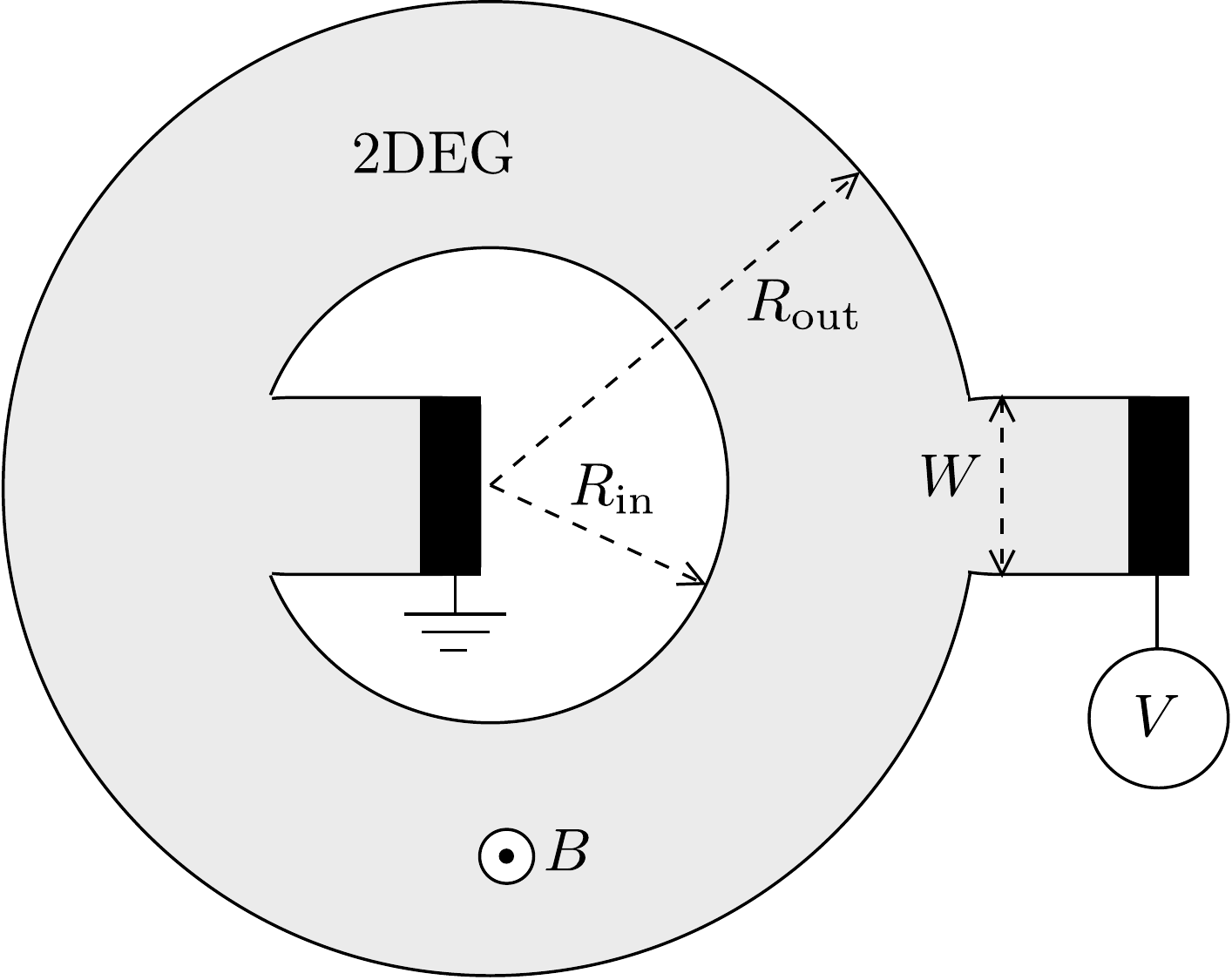}
\caption{Corbino geometry consisting of a ring-shaped two-dimensional electron gas (2DEG, grey) in a perpendicular magnetic field $B$. A pair of electrodes (black), at a voltage difference $V$, is attached to the inner and outer edges of the ring.}
\label{fig:corbino}
\end{figure}

To isolate $G_{\rm bulk}$ from $G_{\rm edge}$ we consider a Corbino geometry, see Fig.\ \ref{fig:corbino}, consisting of a ring-shaped two-dimensional electron gas (2DEG) connected to an electrode at the inner and outer perimeters. The two-terminal conductance $G=G_{\rm bulk}$ is then fully due to the current $I$ through the bulk, without contributions from the currents circulating along the edges.

We assume that the outer electrode, at a voltage $V$ relative to the inner electrode, is connected via an $N$-mode lead to the 2DEG. (For simplicity we do not include spin as a separate degree of freedom.) The reflection amplitudes $r_{mn}$ from mode $n$ to mode $m$ are contained in the $N\times N$ reflection matrix $r$. The conductance $G=I/V$ (in units of the conductance quantum $G_{0}=e^{2}/h$) follows from
\begin{equation}
G=G_{0}\,{\rm Tr}\,(\openone -rr^{\dagger})=G_{0}\sum_{n}(1-R_{n}),\label{Grrelation}
\end{equation}
with $\openone$ the $N\times N$ unit matrix. The reflection eigenvalues $R_{n}\in[0,1]$ are eigenvalues of the Hermitian matrix product $rr^{\dagger}$. Away from the plateau transition all $R_{n}$'s are close to unity and $G\ll e^{2}/h$.

The topological quantum number ${\cal Q}\in\mathbb{Z}$ of the quantum Hall effect in a translationally invariant system is the Chern number of the bands below the Fermi level.\cite{Tho82} An alternative formulation,\cite{Bra09,Ful11} applicable also to a finite disordered system, obtains ${\cal Q}$ as the winding number of the determinant of the reflection matrix,
\begin{equation}
{\cal Q}=\frac{1}{2\pi i} \int_{0}^{2\pi} d\phi\, \frac{d}{d\phi}\ln\det r(\phi),\label{Qdetrphi}
\end{equation}
where $\Phi\equiv\phi\times \hbar/e$ is the magnetic flux enclosed by the inner perimeter of the ring. To define the winding number one needs to vary $\phi$ at constant magnetic field in the 2DEG, so that $r(\phi)$ is $2\pi$-periodic. The conductance \eqref{Grrelation} then oscillates periodically as a function of $\phi$, with amplitude $\Delta G$. (We will examine in the next Section to what extent this applies to the realistic situation of a uniform magnetic field.) 

\begin{figure}[tb]
\includegraphics[width = 0.6\linewidth]{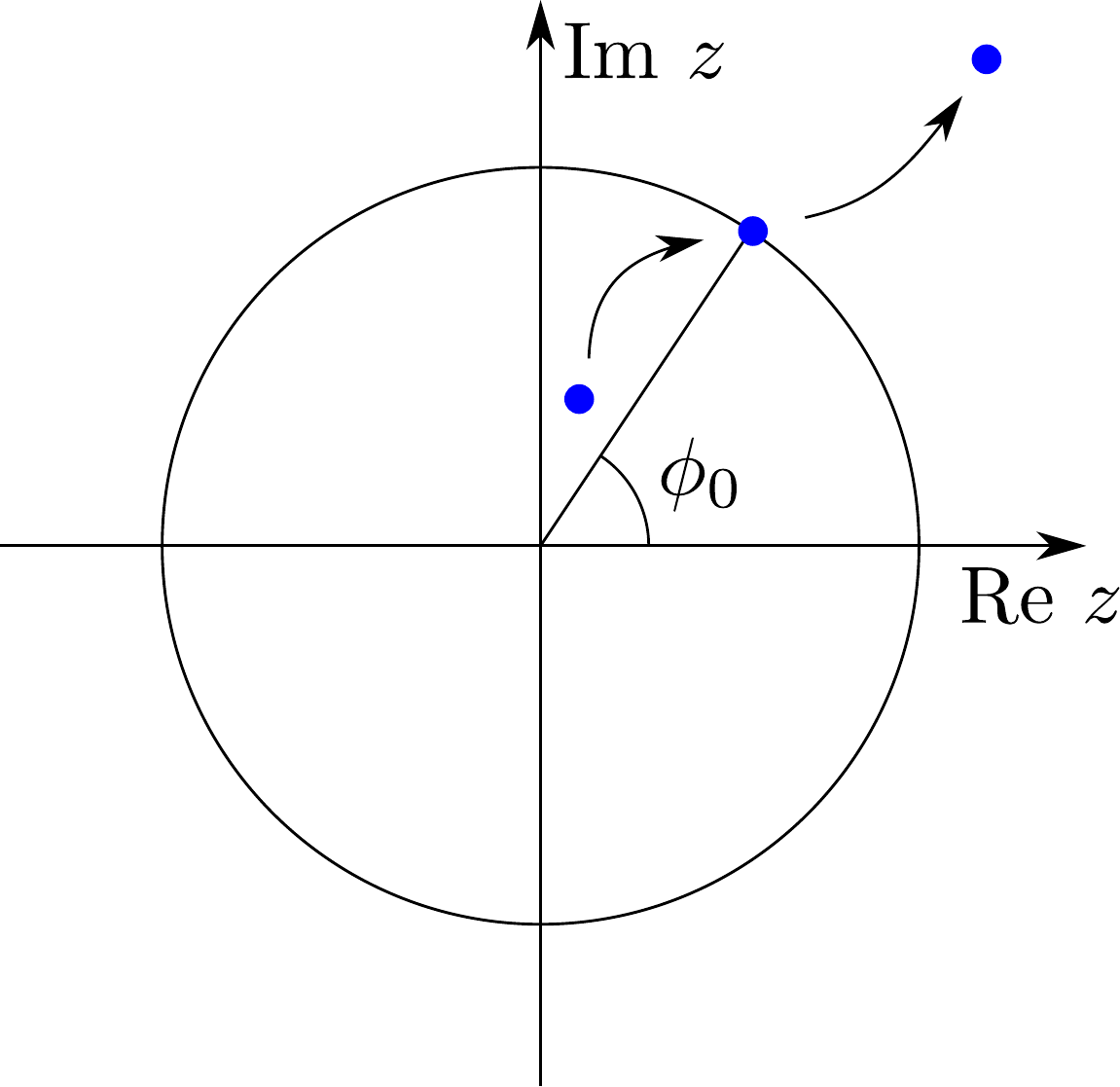}
\caption{Example of a path taken in the complex plane $z=e^{i\phi}$ by a zero (blue dot) of $\det r$, as the system is driven through a phase transition in which ${\cal Q}\rightarrow{\cal Q}-1$. At the transition, the zero crosses the unit circle and a reflection eigenvalue vanishes for some flux $\phi_{0}$. The speed at which the zero crosses the unit circle determines the critical exponent of the quantum Hall phase transition, see Sec.\ \ref{conclude}.}
\label{fig:zero}
\end{figure}

If the $\phi$-dependence of $r$ is continued analytically to arbitrary complex $z=e^{i\phi}$, the winding number
\begin{equation}
{\cal Q}=n_{\rm zero}-n_{\rm pole}\label{Qn0ninfty}
\end{equation}
equals the difference of the number of zeros and poles of $\det r$ inside the unit circle $|z|=1$. A pole may annihilate a zero, but the difference $n_{\rm zero}-n_{\rm pole}$ can only change when a pole or zero crosses the unit circle. Unitarity of the scattering matrix requires $|\det r(\phi)|\leq 1$ for real $\phi$, so poles cannot cross the unit circle. A topological phase transition, corresponding to a change in ${\cal Q}$, is therefore signaled by a zero of $\det r$ crossing the unit circle,\cite{Ful11} see Fig.\ \ref{fig:zero}.

We conclude that at the transition point $\det r(\phi)=0$ for some real value of $\phi$, which implies that (at least) one of the $R_{n}$'s vanishes. If the zeros are well separated, and all other $R_{n}$'s remain close to unity, the conductance oscillation amplitude $\Delta G$ would thus show a peak of $e^{2}/h$ whenever ${\cal Q}$ changes by $\pm 1$. This transport signature of a topological phase transition is the analogue for a 2D system of the 1D signature of Ref.\ \onlinecite{Akh11}.

\section{Numerical simulations in a disordered system}
\label{xp}

We have searched for the predicted conductance resonances in a numerical simulation. The Corbino disc is discretized on a square lattice (lattice constant $a$, nearest-neigbor hopping energy $t$). Electrostatic disorder is introduced as an uncorrelated random on-site potential, drawn uniformly from the interval $(-\epsilon_0, \epsilon_0)$. A uniform perpendicular magnetic field $B$ is introduced by the Peierls substitution, with $f=Ba^{2}e/h$ the flux through a unit cell in units of the flux quantum.

Referring to Fig.\ \ref{fig:corbino} we take parameters $R_{\rm out} = 200\,a$, $R_{\rm in} = 150a$, $W= 60\,a$. For $f = 0.005$ we have a magnetic length $l_m=(\hbar/eB)^{1/2} = 5.6\,a$, sufficiently small that the edge states at opposite edges do not overlap. These parameter values are representative for a $\mu{\rm m}$-size GaAs Corbino disc in a field of $5\,{\rm T}$. The disorder strength was fixed at $\epsilon_0 = 0.2\,t$.

\begin{figure*}[tb]
\includegraphics[width = 0.4\textwidth]{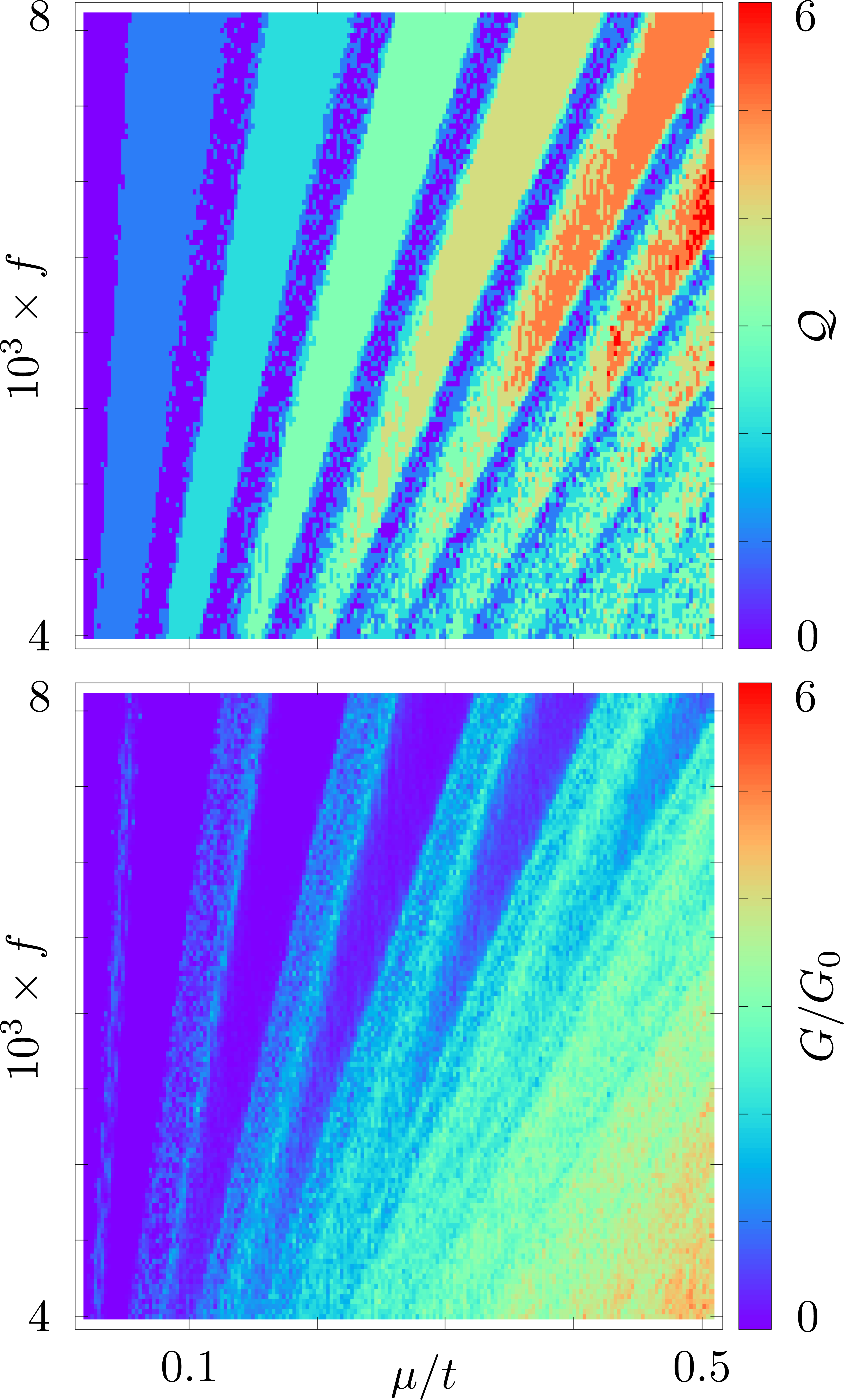}\hfill
\includegraphics[width = 0.4\textwidth]{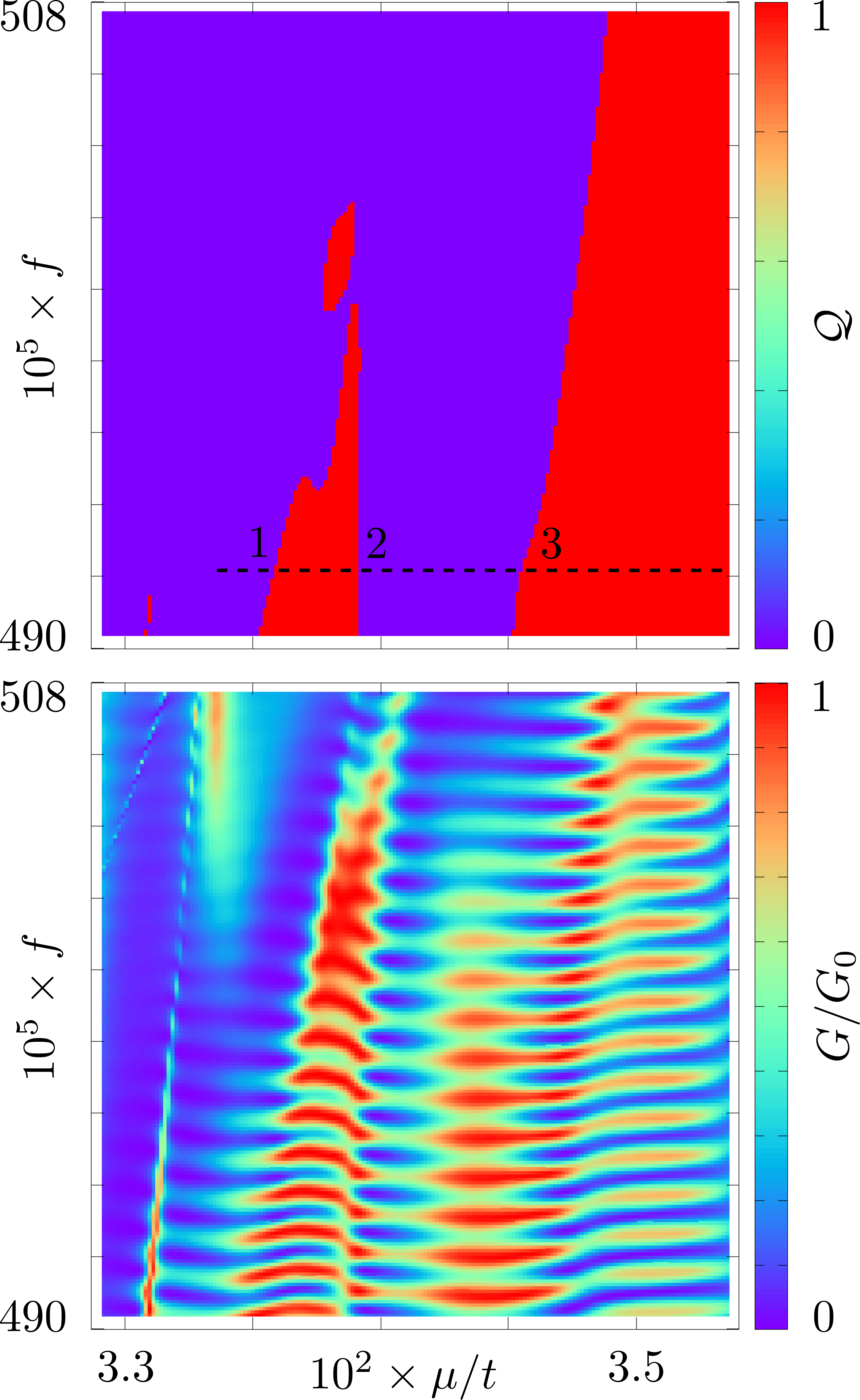}
\caption{Topological quantum number (top panels) and conductance (bottom panels) of a disordered 2DEG in the Corbino disk geometry. The left panels give a broad range of flux $f$ and Fermi energy $\mu$, while the right panels show a close-up of the ${\cal Q}=0\rightarrow 1$ transition. (Note the different color scales for the left and right panels.) The horizontal dashed line indicates the three reentrant phase transitions (see Fig.\ \ref{fig:path}).
}
\label{fig:bev}
\end{figure*}

We calculate the reflection matrix $r$ at energy $\mu$ as a function of the flux $f$ using the recursive Green function technique. This gives the conductance $G$ via Eq.\ \eqref{Grrelation}. The topological quantum number ${\cal Q}$ is calculated from Eq.\ \eqref{Qn0ninfty}, by locating the zeros and poles of $\det r$ in the complex flux plane using the method of Ref.\ \onlinecite{Ful11}. (We summarize this method in App.\ \ref{Appwinding}.) Results are shown in Fig.\ \ref{fig:bev}. Since we are interested in mesoscopic, sample-specific effects, this is data for a single disorder realization, without ensemble averaging.

The left panels show that we can access regions with different topological quantum number by varying flux and Fermi energy. This represents a wide sweep in parameter space, with plateau regions of constant ${\cal Q}=1,2,3,4,5$ separated by transition regions of fluctuating ${\cal Q}$. The conductance $G$, which in the Corbino geometry is the bulk conductance, vanishes on the plateau regions and fluctuates in the transition regions.

The right panels zoom in on the ${\cal Q}=0$ to $1$ transition. The conductance oscillations as a function of the flux are only approximately periodic, because the magnetic field changes uniformly (and not just inside the inner perimeter of the ring). Still, a dominant periodicity $\Delta f=1.1 \cdot 10^5$ can be extracted, corresponding to an $h/e$ flux through a disc of radius $R_{\rm eff}\approx 170\,a$, about halfway between the inner and outer radii of the ring. (For a $\mu{\rm m}$-size Corbino disc this would correspond to a field scale of $\Delta B\approx 10\,{\rm mT}$.)

\begin{figure}[tb]
 \includegraphics[width = 0.4\textwidth]{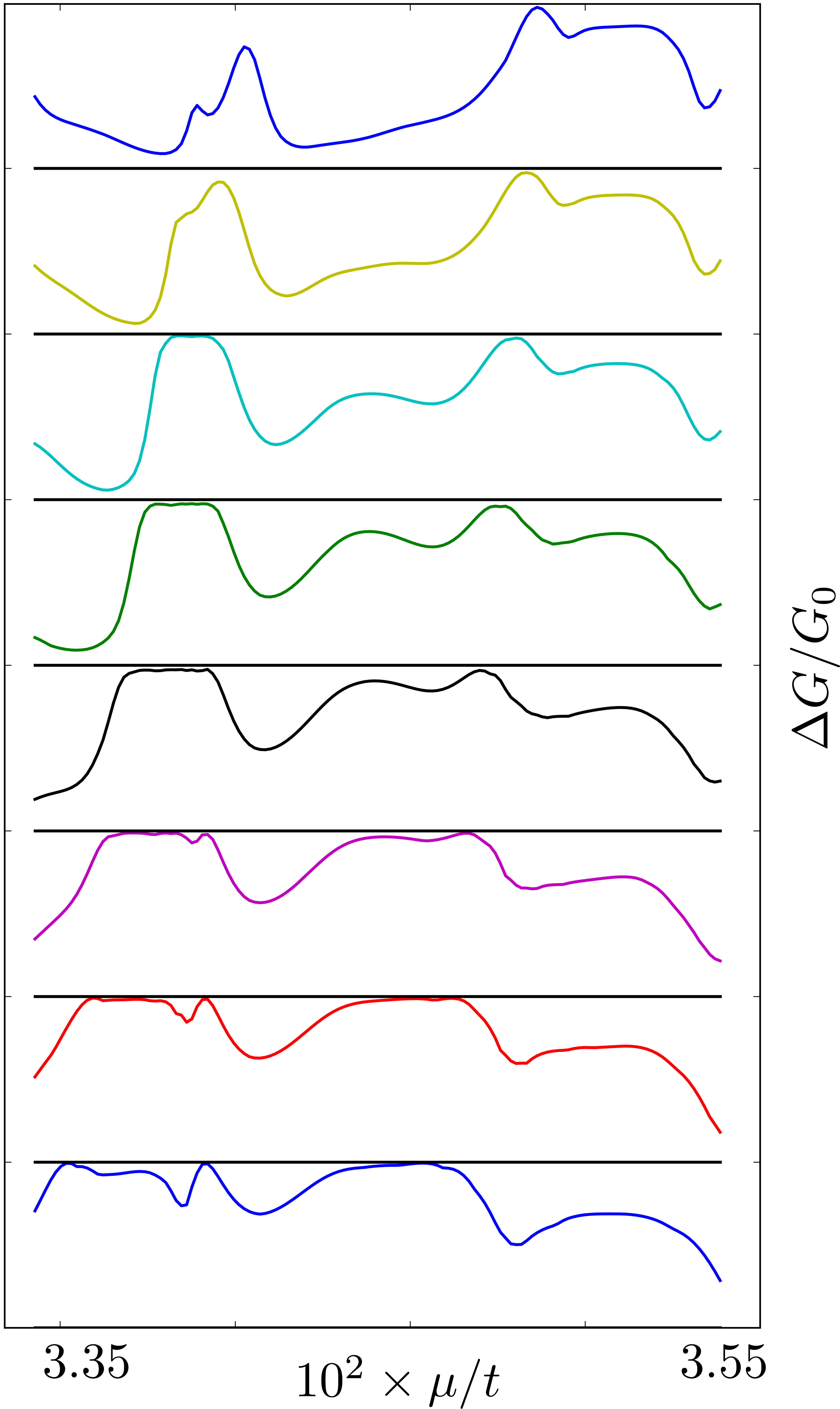}
 \caption{Dependence of the amplitude of the magnetoconductance oscillations on the Fermi energy for different magnetic fields. The $n$-th trace from the bottom shows the $\mu$-dependence of $\Delta G=G_{\rm max}-G_{\rm min}$ in the flux interval $(f_{n},f_{n}+\Delta f)$, with $f_{n}=490\cdot 10^{-5}+2(n-1)\Delta f$ and $\Delta f=1.1 \cdot 10^5$. Successive traces are displaced vertically, with the $G_{0}=e^{2}/h$ peak value indicated by a horizontal black line.}
\label{fig:maxg}
\end{figure}

To search for the predicted signature of the topological phase transition, we calculate the amplitude $\Delta G=G_{\rm max}-G_{\rm min}$ of the conductance oscillations over one period $(f,f+\Delta f)$. This amplitude is plotted in Fig.\ \ref{fig:maxg} as a function of Fermi energy, for different fluxes $f$. The ${\cal Q}=0\rightarrow 1$ transition should give at least one $e^{2}/h$ peak value of $\Delta G$, which is indeed what is observed. For the larger field values there is only a single $e^{2}/h$ peak, while for the lower field values three peaks develop, consistent with the ${\cal Q}=0\rightarrow 1\rightarrow 0\rightarrow 1$ reentrant transition seen in Fig.\ \ref{fig:bev} (upper right panel, dashed line).  

\section{Discussion and relation to the critical exponent}
\label{conclude}

We have shown that the sample-specific conductance fluctuations at the quantum Hall plateau transition contain a universal feature of a topological origin: The magnetoconductance oscillations in a Corbino disc geometry have an $e^{2}/h$ resonant amplitude whenever the topological quantum number ${\cal Q}$ is incremented by $\pm 1$. The strict $h/e$ flux periodicity of the magnetoconductance is broken in a realistic setting by the penetration of the flux through the conducting region, but we have shown by numerical simulations that the resonances are still clearly observable.

The theory presented here may motivate an experimental search for the reentrant phase transition that we observed in our simulations. Existing experiments\cite{Hoh02} in a GaAs Corbino disc show a maximal amplitude of about $0.1\,e^{2}/h$ in the ${\cal Q}=1\rightarrow 2$ plateau transition, an order of magnitude below the predicted value. This may partly be due to limited phase coherence, and partly to the self-averaging effect of overlapping resonances. In the experimental geometry the entire inner and outer perimeters of the Corbino disc are contacted to current source and drain, while our geometry (Fig.\ \ref{fig:corbino}) has two narrow contacts --- which helps to isolate the resonances and make them more easily observable.

From a theoretical perspective, the evolution of zeros of the reflection matrix determinant, see Figs.\ \ref{fig:zero} and \ref{fig:path}, provides a new way to analyse the plateau transition in a disordered mesoscopic system. In particular, the critical exponent $\nu$ of the quantum Hall phase transition can be extracted from the speed at which a zero $z_{0}$ of $\det r(z)$ crosses the unit circle in the complex flux plane,
\begin{equation}
\lim_{L\rightarrow\infty}\frac{W}{L}\,\ln |z_{0}|={\rm const}\times (\mu-\mu_{c})W^{1/\nu}+{\cal O}(\mu-\mu_{c})^{2}.\label{scaling}
\end{equation}
Here $\mu_{c}$ is the value of the Fermi energy $\mu$ at the phase transition, where $z_{0}=e^{i\phi}$ for some real flux $\phi$. The Corbino disc has inner perimeter of radius $L$ and outer perimeter of radius $L+W$, with the limit $L\rightarrow\infty$ taken at constant $W$.

We note the following conceptual difference with the usual MacKinnon-Kramer scaling:\cite{Mac81,Sle09} The MacKinnon-Kramer scaling variable is a Lyapunov exponent, which is a nonnegative quantity. Our scaling variable $\ln |z_{0}|$ changes sign at the phase transition, so it contains information on which side of the transition one is located. 

In App.\ \ref{app_scaling} we present a calculation of the critical exponent along these lines. It remains to be seen whether this alternative numerical method has advantages over the conventional method based on MacKinnon-Kramer scaling. It is also still an open question whether the evolution of the zeros of $\det r$ is more accessible to analytical methods than the evolution of Lyapunov exponents.

\acknowledgments

The numerical calculations were performed using the {\sc kwant} package developed by A. R. Akhmerov, C. W. Groth, X. Waintal, and M. Wimmer. Our research was supported by the Dutch Science Foundation NWO/FOM and by an ERC Advanced Investigator Grant.

\appendix

\section{Calculation of the topological quantum number}
\label{Appwinding}

We summarize the method used to calculate the topological quantum number ${\cal Q}$ of the Corbino disc, following Ref.\ \onlinecite{Ful11}. We insert a flux tube $\Phi=\phi\hbar/e$ at the center, without changing the magnetic field in the 2DEG, and obtain ${\cal Q}$ as the winding number of $\det r$. The integration \eqref{Qdetrphi} over $\phi$ can be avoided by going to the complex $z=e^{i\phi}$ plane and counting the number of zeros and poles of $\det r$ in the unit circle. The difference $n_{\rm zero}-n_{\rm pole}$ then directly gives ${\cal Q}$.

\subsection{Analytic continuation to complex flux}
\label{complexflux}

\begin{figure}[tb]
\includegraphics[width = 0.7\linewidth]{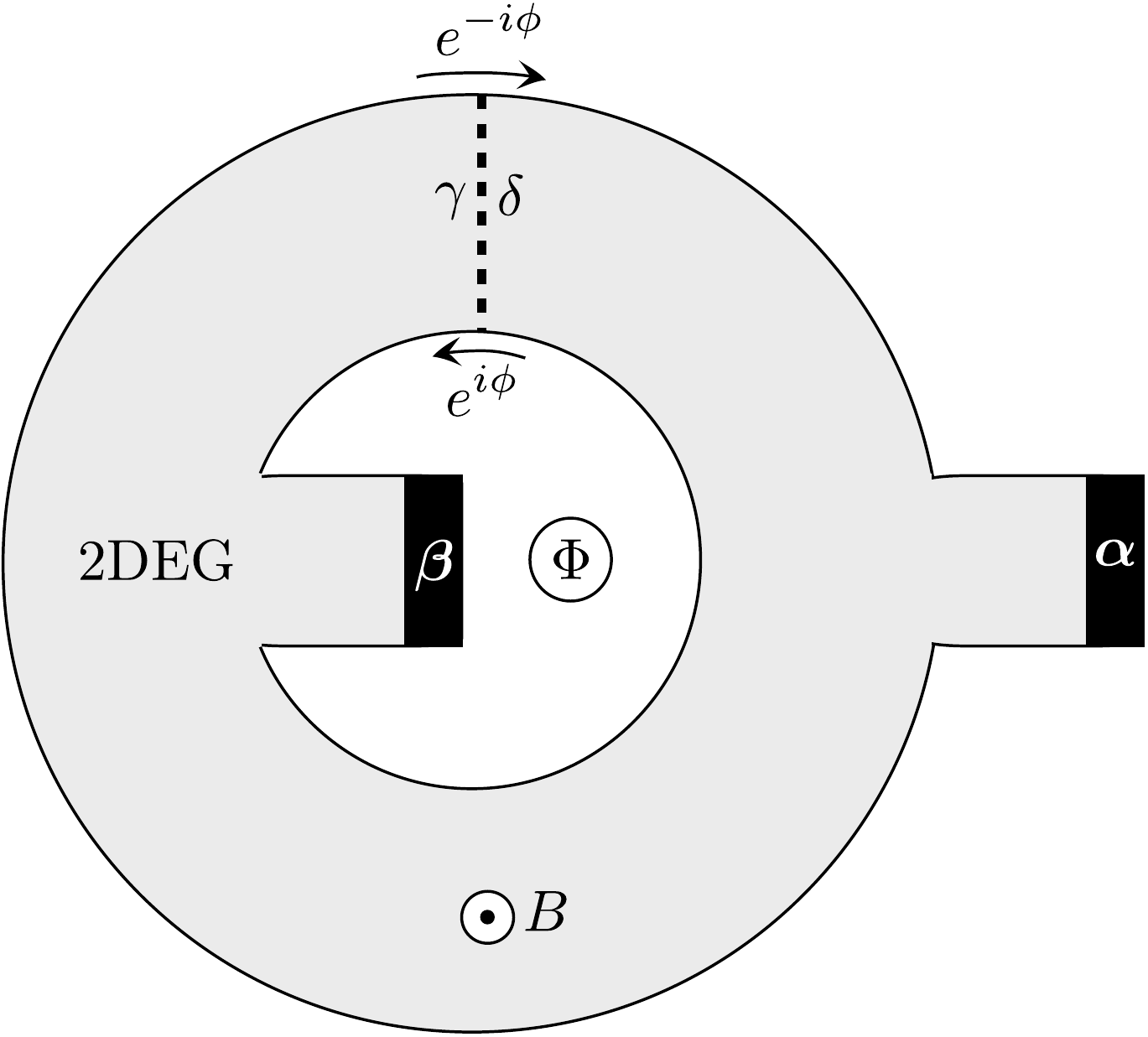}
\caption{Corbino disc in a uniform magnetic field $B$, containing additionally a flux tube $\Phi$ inside the inner perimeter. We choose a gauge such that the phase increment $\phi=e\Phi/\hbar$ due to the flux tube is accumulated upon crossing a cut indicated by the dashed line. Virtual leads $\gamma,\delta$ are attached at the two sides of the cut. Together with the two physical leads $\alpha,\beta$ they define the four-terminal scattering matrix \eqref{eq:scatter}. The reflection matrix $r$ of the original two-terminal system is obtained from Eq.\ \eqref{eq:rphi}, in a formulation which can be analytically continued to arbitrary complex $z=e^{i\phi}$.
}
\label{fig_cut}
\end{figure}

\begin{figure}[tb]
 \includegraphics[width = 0.9\linewidth]{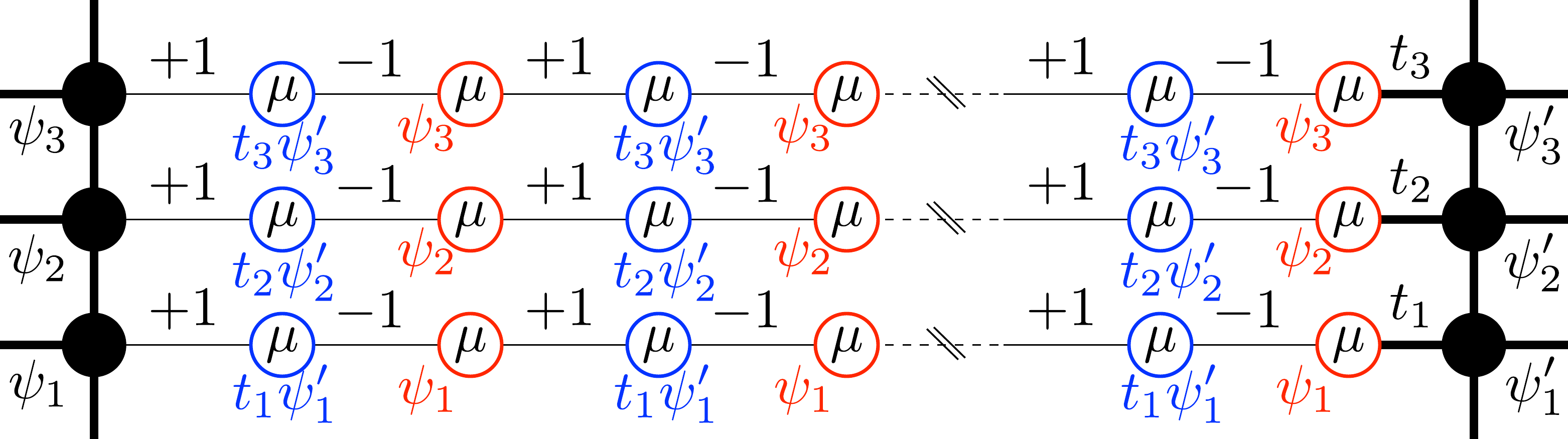}
 \caption{Construction of a virtual lead, connecting two sides of a tight-binding lattice. The virtual lead consists of parallel 1D chains with on-site energy $\mu$ and hopping amplitudes alternating between $+1$ and $-1$. The tight-binding equations enforce that every other site in the chain (colored red or blue) has the same wave amplitude at energy $\mu$. The wave amplitudes $\psi_{n}$, $\psi'_{n}$ at the left and right end of the $n$-th chain are therefore unaffected by the insertion of the virtual lead. We use this device in Fig.\ \ref{fig_cut} to convert a two-terminal geometry into a four-terminal geometry, by cutting the lead at the center and connecting the ends to terminals $\gamma,\delta$.
 \label{fig:vlead}
 }
\end{figure}

To perform the analytic continuation to complex flux we cut the disc as indicated in Fig.\ \ref{fig_cut}. We attach virtual leads (see Fig.\ \ref{fig:vlead}) at the two sides of the cut and construct the four-terminal unitary scattering matrix
\begin{equation}\label{eq:scatter}
 S = 
  \begin{pmatrix}
    r_{\alpha\alpha} & t_{\alpha\beta} & t_{\alpha\gamma} & t_{\alpha\delta} \\
    t_{\beta\alpha} & r_{\beta\beta} & t_{\beta\gamma} & t_{\beta\delta} \\
    t_{\gamma\alpha} & t_{\gamma\beta} & r_{\gamma\gamma} & t_{\gamma\delta} \\
    t_{\delta\alpha} & t_{\delta\beta} & t_{\delta\gamma} & r_{\delta\delta}
  \end{pmatrix}.
\end{equation}
We choose a gauge such that the entire phase increment $\phi$ from the flux tube is accumulated at the cut, between terminals $\gamma$ and $\delta$. The reflection matrix $r$ of the original two-terminal system (from terminal $\alpha$ back to $\alpha$) is then obtained from
\begin{align}\label{eq:rphi}
   r={}& r_{\alpha\alpha} - \\\nonumber
  &  \begin{pmatrix} 
    t_{\alpha\gamma} & t_{\alpha\delta} 
  \end{pmatrix}
  \begin{pmatrix}
    r_{\gamma\gamma} & t_{\gamma\delta}-z \\
    t_{\delta\gamma}- 1/z & r_{\delta\delta}
  \end{pmatrix}^{-1}
  \begin{pmatrix}
    t_{\gamma\alpha} \\
    t_{\delta\alpha}
  \end{pmatrix}.
\end{align}
This expression is in a form suitable for analytic continuation to arbitrary complex $z=e^{i\phi}$.

\subsection{Evolution of zeros and poles of $\bm{\det r}$}

The determinant of the reflection matrix \eqref{eq:rphi} is computed as a function of the complex variable $z$ by means of the relation
\begin{equation}
 \det (D - CA^{-1}B) = \det \begin{pmatrix}
                             A & B \\
			     C & D
                            \end{pmatrix} [ \det A]^{-1}.
\end{equation}
This results in the expression
\begin{align}
\det r={}& \det \begin{pmatrix}
       r_{\gamma\gamma} & t_{\gamma\delta} - z & t_{\gamma\alpha} \\
       t_{\delta\gamma} - 1/z & r_{\delta\delta} & t_{\delta\alpha} \\
       t_{\alpha\gamma} & t_{\alpha\delta} & r_{\alpha\alpha}
      \end{pmatrix}\nonumber\\
      &\times\left[ \det \begin{pmatrix}
       r_{\gamma\gamma} & t_{\gamma\delta} - z \\
       t_{\delta\gamma} - 1/z & r_{\delta\delta}
      \end{pmatrix}\right]^{-1}.\label{detrratio}
\end{align}

\begin{figure}[tb]
 \includegraphics[width = 0.7\linewidth]{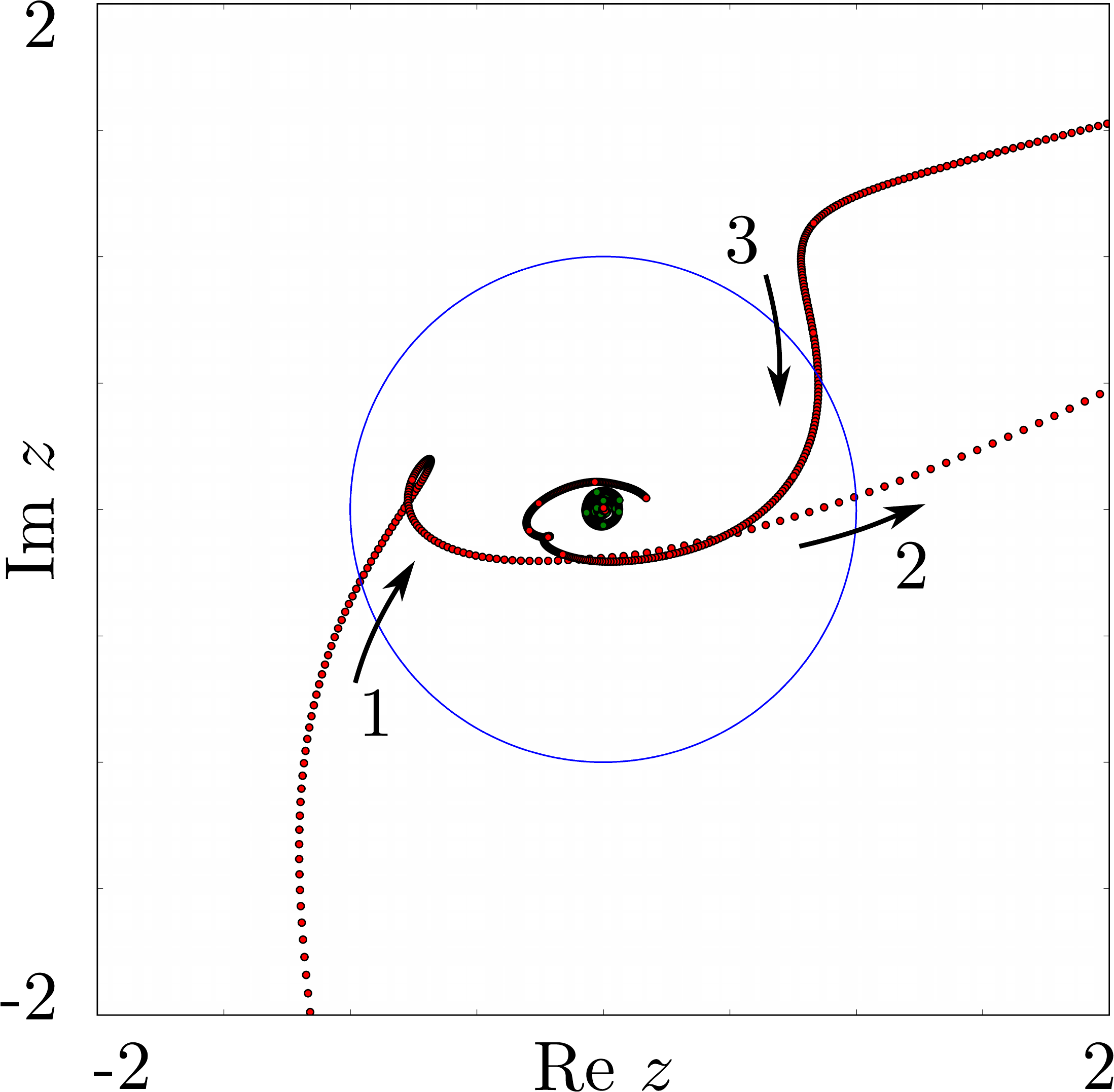}
 \caption{Positions of the zeros (red) and poles (green, clustered near the origin) of $\det r$, along the path indicated by the dashed line in Fig.\ \ref{fig:bev}. Arrows point in the direction of increasing Fermi energy. The numbers $1,2,3$ indicate the three phase transitions in which the topological quantum number ${\cal Q}$ changes first from $0$ to $1$, then from $1$ back to $0$, and once again from $0$ to $1$.}
\label{fig:path}
\end{figure}

The zeros and poles of $\det r$ are therefore those values of $z$ for which, respectively, the determinant in the numerator or denominator of Eq.\ \eqref{detrratio} vanishes. By equating each of these determinants to zero we obtain two generalized eigenvalue problems, with the zeros given by
\begin{equation}
 \begin{pmatrix}
        r_{\gamma\gamma} & t_{\gamma\delta} & t_{\gamma\alpha} \\
	-1               & 0                & 0                \\
	t_{\alpha\gamma} & t_{\alpha\delta} & r_{\alpha\alpha}
        \end{pmatrix} \begin{pmatrix}
\psi_1 \\
\psi_2\\
\psi_3
\end{pmatrix} = -z
\begin{pmatrix}
 0 & -1 & 0 \\
 t_{\delta\gamma} & r_{\delta\delta} & t_{\delta\alpha} \\
 0 & 0 & 0
\end{pmatrix} \begin{pmatrix}
\psi_1 \\
\psi_2\\
\psi_3
\end{pmatrix},
\end{equation}
and the poles by
\begin{equation}
 \begin{pmatrix}
  r_{\gamma\gamma} & t_{\gamma\delta} \\
  -1               & 0
 \end{pmatrix} \begin{pmatrix}
\psi_1 \\
\psi_2
\end{pmatrix} = z
\begin{pmatrix}
 0 & 1 \\
 -t_{\delta\gamma} & -r_{\delta\delta}
\end{pmatrix} \begin{pmatrix}
\psi_1 \\
\psi_2
\end{pmatrix}.
\end{equation}
The numerical solution of generalized eigenvalue problems is quick and accurate. As an example, we show in Fig.\ \ref{fig:path} the motion of the zeros and poles along the reentrant phase transition of Fig.\ \ref{fig:bev}.

\section{Calculation of the critical exponent}
\label{app_scaling}

In Sec.\ \ref{conclude} we outlined how the critical exponent of the quantum Hall phase transition can be extracted from the speed at which a zero $z_{0}$ of $\det r(z)$ crosses the unit circle in the complex $z=e^{i\phi}$ plane. Here we demonstrate that this approach produces results consistent with earlier calculations based on the scaling of Lyapunov exponents.\cite{Sle09}

To make contact with those earlier calculations, we use the same Chalker-Coddington network model \cite{Cha88,Kra05} (rather than the tight-binding model used in the main text). The parameter that controls the plateau transition in the network model is the mixing angle $\alpha$ of the scattering phase shifts at the nodes of the network. The transition is at $\alpha_{c}=\pi/4$. Disorder is introduced by means of random, uncorrelated phase shifts on the links between nodes, sampled uniformly from $[0,2\pi)$. 

\begin{figure}[tb]
\includegraphics[width = 0.8\linewidth]{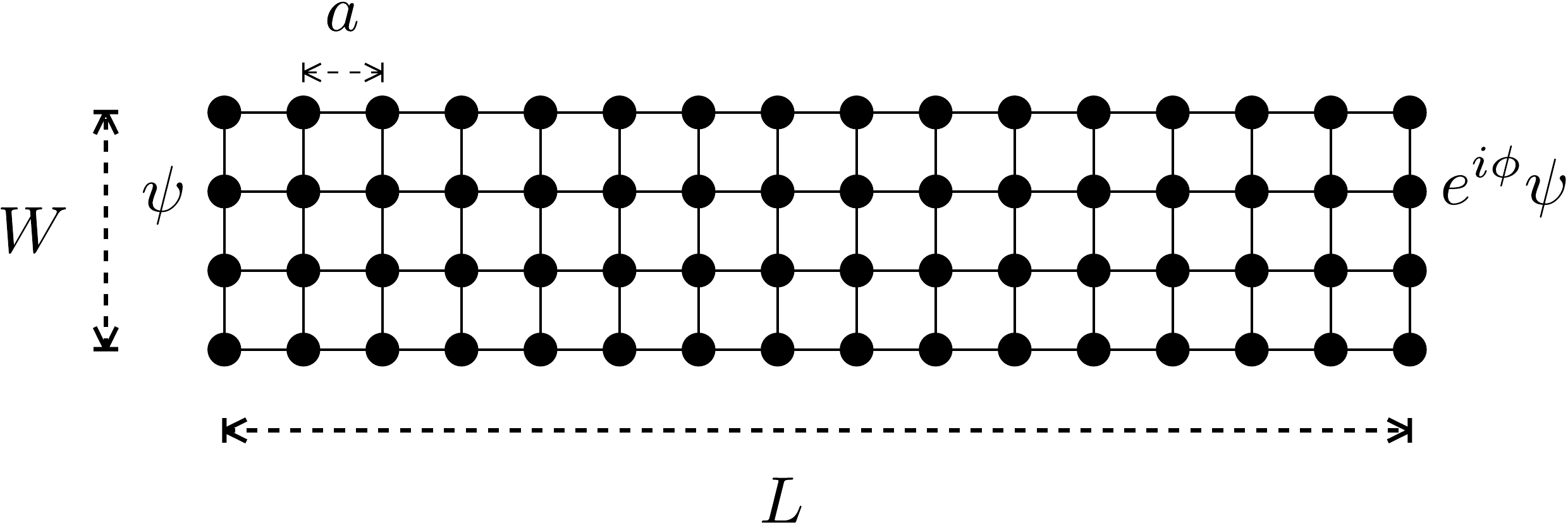}
\caption{Geometry of the network model used to calculate the critical exponent. This rolled up strip is topologically equivalent to the Corbino geometry used in the main text.
}
\label{fig_network}
\end{figure}

The network has a strip geometry, of longitudinal dimension $L$ and transverse dimension $W$ (see Fig.\ \ref{fig_network}). The longitudinal dimension has twisted boundary conditions, $\psi\mapsto e^{i\phi}\psi$ upon translation over $L$. The transverse dimension is connected to a source and drain reservoir. This geometry is equivalent to a Corbino disc, enclosing a flux $\Phi=\phi\hbar/e$ and with source and drain contacts extending along the entire inner and outer perimeter. 

\begin{figure}[tb]
\includegraphics[width = 0.9\linewidth]{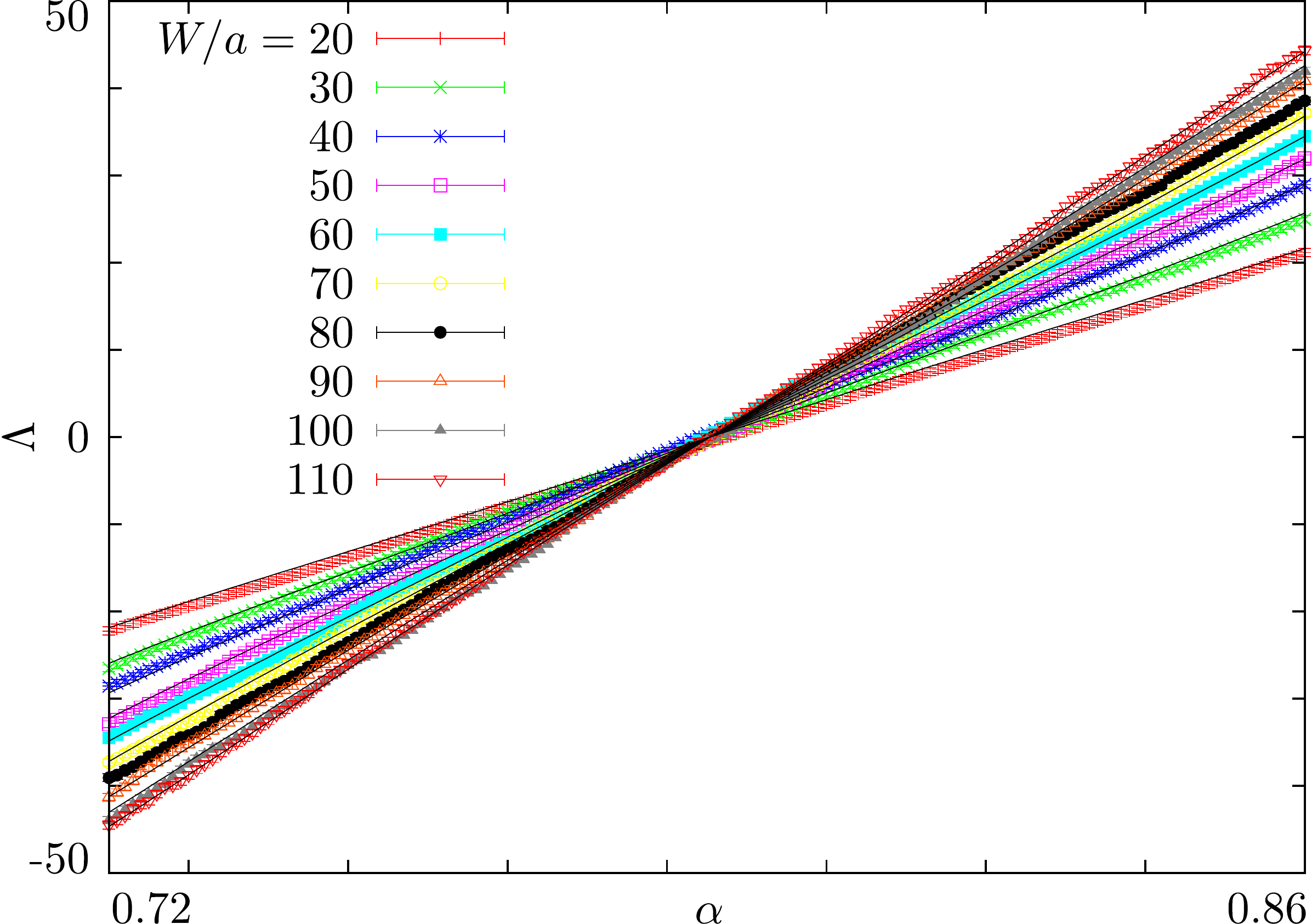}
\caption{Scaling variable as obtained numerically (data points) and as fitted to Eq.\ \eqref{scalingfit} (black lines).
}
\label{fig:exponent}
\end{figure}

Taking the self-averaging limit $L\rightarrow\infty$ of Eq.\ \eqref{scaling} is impractical, but sufficient convergence is reached for an aspect ratio $L/W=5$ and an average of $\ln |z_{0}|$ over 1000 disorder realizations. Results for the $W$ and $\alpha$-dependence of the scaling parameter $\Lambda=(W/L)\langle\,\ln |z_{0}|\rangle$ are shown in Fig. \ref{fig:exponent}.

We extract the critical exponent $\nu$ using the procedure of Ref.\ \onlinecite{Sle09}, by fitting the data to the scaling law
\begin{equation}
\Lambda(W,\alpha)=\sum_{p=1}^{n}c_{p}W^{p/\nu}\left[\sum_{q=1}^{n'}c'_{q}(\alpha-\alpha_{c})^{q}\right]^{p}.\label{scalingfit}
\end{equation}
We have checked that corrections to scaling are insignificant in our parameter range. The symmetry of the network model (a time-reversal operation followed by translation over half a unit cell) implies that only odd powers of $\alpha-\alpha_c$ appear in the series expansion ($c_p=c'_p=0$ for $p$ even). We truncated the expansions such as to keep terms of order $(\alpha - \alpha_c)^5$ and lower. The result $\nu = 2.56 \pm 0.03$ is consistent with the value $\nu=2.59$ obtained in Ref.\ \onlinecite{Sle09}.

\end{document}